\documentclass[twoside,slac_one]{revtex4}
\usepackage{graphicx}
\usepackage{fancyhdr}
\usepackage{atlasphysics}
\usepackage{subfigure}
\usepackage{amsmath} 
\usepackage{bm}
\usepackage{amsxtra}
\usepackage{amssymb}
\usepackage{amsthm}
\usepackage{latexsym}
\usepackage{lscape}
\usepackage{epsfig}

\begin{document}

\title{Measurement of multi-jet cross sections at ATLAS}

%

\author{M. Tamsett on behalf of the ATLAS Collaboration}
\affiliation{Department of Physics, Louisiana Tech University, Ruston, LA, USA.}

\begin{abstract}
Inclusive multi-jet production is studied using the ATLAS detector for proton-proton collisions with a 
center-of-mass energy of 7~$\TeV$.  The data sample corresponds to an integrated luminosity of 
2.43~pb$^{-1}$, using the first proton-proton data collected by the ATLAS detector in 2010.  Results 
on multi-jet cross sections are presented and compared to both 
leading-order plus parton-shower Monte Carlo predictions and next-to-leading-order QCD calculations.
\end{abstract}

\maketitle

\thispagestyle{fancy}


\newcommand{\squishlist}{
 \begin{list}{$\bullet$}
  { \setlength{\itemsep}{0pt}
     \setlength{\parsep}{3pt}
     \setlength{\topsep}{3pt}
     \setlength{\partopsep}{0pt}
     \setlength{\leftmargin}{1.5em}
     \setlength{\labelwidth}{1em}
     \setlength{\labelsep}{0.5em} } }

\newcommand{\squishlisttwo}{
 \begin{list}{$\bullet$}
  { \setlength{\itemsep}{0pt}
     \setlength{\parsep}{0pt}
    \setlength{\topsep}{0pt}
    \setlength{\partopsep}{0pt}
    \setlength{\leftmargin}{2em}
    \setlength{\labelwidth}{1.5em}
    \setlength{\labelsep}{0.5em} } }

\newcommand{\squishend}{
  \end{list}  }

\section{Introduction}

This document represents a summary of the analysis and results presented in \cite{multijetpaper}.

\vspace{5mm}

At hadron colliders, events containing multiple jets in the final state are prolific and provide a fertile 
testing ground for the theory of the strong interaction, quantum chromodynamics (QCD).  

In this paper, a first study is performed of multi-jet events produced from proton-proton collisions at \hbox{7\,TeV} 
center-of-mass energy, using the ATLAS detector at the Large Hadron Collider (LHC) at CERN.  
The data sample used for this analysis was collected between April 10 and August 30 of 2010 
and represents a total integrated luminosity of 2.43~pb$^{-1}$.  

The two primary motivations for the multi-jet study in this paper are to evaluate 
how robust leading-order (LO) QCD calculations are in representing the high jet 
multiplicity events, relevant as backgrounds in high-energy searches, and to 
test next-to-leading-order perturbative QCD (NLO pQCD) calculations.  

\section{The ATLAS Detector}\label{sec:detector}
The ATLAS experiment consists of an approximately 45-meter long, 25-meter diameter 
cylindrically 
shaped detector centered on the proton-proton interaction point.  A detailed 
description of the ATLAS experiment can be found elsewhere~\cite{atlas_detector_paper}. 

High-energy 
particles produced in collisions initially pass through an inner tracking system embedded in a \hbox{$2\,$T} solenoidal magnetic field.  
Just outside the inner tracker package is a 
system of liquid argon and scintillating tile calorimeters used for the measurement of particle 
energies.  A liquid-argon/lead electromagnetic calorimeter covers the 
pseudorapidity\footnote{ATLAS uses a right-handed coordinate system with its origin at the nominal interaction point (IP) in the centre of the detector and the $z$-axis along the beam pipe. The $x$-axis points from the IP to the centre of the LHC ring, and the $y$ axis points upward. Cylindrical coordinates $(r,\phi)$ are used in the transverse plane, $\phi$ being the azimuthal angle around the beam pipe. The pseudorapidity is defined in terms of the polar angle $\theta$ as $\eta=-\ln\tan(\theta/2)$. The rapidity is defined as $y = 0.5 \times {\rm ln}[(E+p_z)/(E-p_z)]$, where $E$ denotes the energy and $p_z$ is the component of the momentum along the beam direction. For massless objects, the rapidity and pseudorapidity are equivalent.}  range of $|\eta| < 3.2$.  
This calorimeter is complemented by hadronic calorimeters, built 
using scintillating tiles and iron for $|\eta| < 1.7$ and liquid argon and copper in the 
end-cap ($1.5 < |\eta| < 3.2$).  
Forward calorimeters extend the coverage for ATLAS to $|\eta| = 4.9$. 
The calorimeters are the primary detectors used to reconstruct the jet energy in this analysis.

The ATLAS trigger system employs three trigger levels, of which only the hardware-based first level 
trigger is used in this analysis. Events are selected using the calorimeter-based jet trigger. 
The first level jet trigger~\cite{atlas_trigger08} uses coarse detector information 
to identify areas in the calorimeter where energy deposits above a certain threshold occur. A 
simplified jet finding algorithm based on a sliding window of size 
$\Delta\phi\times\Delta\eta=0.8\times 0.8$ is used to identify 
these areas. This algorithm uses coarse calorimeter towers with a 
granularity of $\Delta\phi\times\Delta\eta=0.2\times 0.2$ as inputs.

\section{Cross Section Definitions and Kinematics}\label{sec:definitions}
The anti-$k_t$ algorithm~\cite{Cacciari_antikt,Fastjet} with full four-momentum recombination 
is used to identify jets. 
For high multiplicity studies, namely up to six jets in an event,
the resolution parameter in the jet reconstruction 
is fixed to $R = 0.4$ to contend with the limited phase space and to reduce the impact of the 
underlying event~\cite{UE} in the jet energy determination. 
For testing NLO pQCD calculations, 
where the study focuses 
on three-jet events, 
a resolution parameter of $R = 0.6$ is preferred, 
since a larger value of $R$ is less sensitive to theoretical 
scale uncertainties.  

Jet measurements are corrected for all experimental effects and refer to the particle-level final 
state. 
At the particle level, jets are built using all final-state particles with a proper 
lifetime longer than 10~ps, including muons and neutrinos from hadronic decays. These corrections
are described in Section~\ref{sec:unfolding}. The NLO pQCD calculation is 
not interfaced
to a Monte Carlo (MC) simulation with hadronization and other non-perturbative effects.  The correction for non-perturbative effects applied to the NLO pQCD calculation is described in Section~\ref{sec:theory}.

Cross sections are calculated in bins of inclusive jet multiplicity, 
meaning that an event is recorded in a jet multiplicity bin if it contains a number of jets that is equal to or greater than that multiplicity.  
 
Inclusive multiplicity bins are used because they are stable
in the pQCD fixed-order calculation, unlike exclusive bins. 
Only jets with $\pt\geq60\GeV$ and $|y|\leq2.8$ are counted in the analysis.  These cuts are chosen to ensure that the jets are reconstructed with high efficiency.  
The leading jet is further required
to have $\pt\geq80\GeV$ to stabilize the NLO pQCD calculations~\cite{Frixione}. 

\section{Theoretical Predictions}\label{sec:theory}

Measurements are compared to pQCD calculations at leading order and next-to-leading order.
For the leading-order analysis, ALPGEN~\cite{Mangano:2002ea} is used to 
generate events with up to six partons in the 
final state using the leading-order set of proton pdfs
CTEQ6L1~\cite{Pumplin:2002vw}. 
ALPGEN is interfaced to 
PYTHIA 6.421~\cite{Sjostrand:2007gs,Sjostrand:2000wi} and, alternatively, to
HERWIG/JIMMY \cite{Corcella:2000bw,Corcella:2002jc,Butterworth:1993ig,Butterworth:1996zw} to 
sum leading logarithms to all orders in the parton-shower approximation and to include 
non-perturbative effects such as hadronization and the underlying event.
The ATLAS generator tunes from 2009 (MC09$^{\prime}$)
~\cite{ATL-PHYS-PUB-2010-002} and from 2010 (AUET1)~\cite{ATL-PHYS-PUB-2010-014} are
used. 
Additional tunes are
investigated to assess the impact of the underlying event and parton-shower tuning. 
With comparable underlying event tunes and ALPGEN
parameters, the comparison between ALPGEN+PYTHIA an ALPGEN+HERWIG/JIMMY uncovers 
differences that may arise from different parton-shower implementations and 
hadronization models. 

SHERPA~\cite{Sherpa} with its default parameters and renormalization scale scheme from version 1.2.3 is also used to generate events with up to six partons in the final state.  This provides an independent matrix element calculation with a different matrix element and parton shower matching scheme.  

The PYTHIA 6.421~\cite{Sjostrand:2007gs,Sjostrand:2000wi} event generator is included to 
study the limitations in a $2\rightarrow 2$ calculation. 
The MRST 2007 modified leading order~\cite{Sherstnev:2008dm,Martin:2009iq}
pdfs interfaced with the AMBT1\cite{ATL-PHYS-PUB-2010-002} 
generator tune is used in the sample generation. 

The generated particles are passed through a full 
simulation of the ATLAS detector and trigger~\cite{SimJINST} based on GEANT4~\cite{geant} 
to account for detector effects. 
Additional proton-proton collisions are added to the hard scatter in the simulation process
to reproduce realistic LHC running conditions. 
Events and jets are selected using the same criteria in data and Monte Carlo simulations.

For the NLO pQCD study, the calculation implemented in NLOJet++~\cite{NLOJet} 
is used. 
The renormalization and factorization scales were varied independently by a factor of two in order to estimate the impact of higher order terms not included in the calculation.  An additional requirement that the ratio of the renormalization and factorization scales did not differ by more than a factor of two was imposed.
Two NLO pdf sets 
CTEQ 6.6~\cite{Nadolsky:2008zw} and MSTW 2008 nlo~\cite{Martin:2009iq}
were used for calculating the central values.
The 90\% confidence-limit
error sets are used in the evaluation of the pdf uncertainties.
The uncertainty
in the measurement due to the uncertainty in the value of $\alpha_S$ is calculated by varying the value of 
$\alpha_S$ by $\pm0.002$ for each pdf set. 

The NLOJet++ program is a matrix element calculation, and therefore it lacks a 
parton-shower interface and
does not account for non-perturbative effects. To compare to particle-level
jet cross sections, supplementary calculations are required. PYTHIA and 
HERWIG++~\cite{Bahr:2008pv}
are used to generate samples without underlying event. Jets in these samples are reconstructed
from partons after the parton shower, and observables are compared at the particle-level in the standard HERWIG++ and PYTHIA samples. A 
multiplicative correction is calculated.
This correction factor takes the NLO pQCD calculations to the particle level. 
The correction
obtained using the PYTHIA AMBT1 sample is taken as the default 
value for the analysis, and the 
systematic uncertainty is estimated from the
maximum spread compared to the results from the other models. The size of this
correction is less than 5\% in all observables studied in the NLO pQCD analysis. 
The total uncertainty quoted on the NLO pQCD calculations comes from the 
quadrature sum of the uncertainties from the renormalization and factorization scales, the proton pdfs, $\alpha_S$ 
and the non-perturbative corrections. 

\section{Event Selection and Reconstruction}\label{sec:event-sel}

A set of ATLAS first level (level-1) multi-jet triggers is used to select events for the analysis. 
Multi-jet triggers require several jets reconstructed with a level-1 sliding window algorithm.  All multi-jet triggers are symmetric. Only two-jet and three-jet triggers were needed for the analysis.

The primary vertex or vertices are found using tracks that originate in the beam collision 
spot\cite{ATLAS-CONF-2010-027}, satisfy quality criteria~\cite{ATLAS-CONF-2010-069} 
and have transverse momentum above $150\MeV$. 

Topological calorimeter clusters evaluated at the electromagnetic scale~\cite{:2010wv} 
are used as inputs to the jet finding algorithm. 
The anti-$k_t$ algorithm~\cite{Cacciari_antikt} with resolution parameters 
$R=0.4$ and $R=0.6$ and full four-momentum recombination 
is used to reconstruct jets from clusters.  
The jet reconstruction is fully efficient in the Monte Carlo 
simulation for jets with transverse momentum above $30\GeV$. 
The ATLAS Monte Carlo simulation compares well with the jet reconstruction 
efficiency measured with data~\cite{:2010wv}. 

Jets reconstructed at the electromagnetic scale are measured to have an energy which 
is lower than the true energy of interacting particles within the jet. This is caused by 
the different response of the calorimeter to hadrons and electrons and by energy lost in 
dead material, deposited outside the jet cone or excluded by the cell clustering algorithm. 
A Monte Carlo-based calibration 
that corrects for these effects as a function of $\pt$ and $y$ is used to obtain 
jets with the correct energy scale~\cite{atlas_jes_study}. 

Jets considered in the analysis are selected using the following kinematic and data quality
selection criteria:
\squishlist
\item An event must contain at least one jet with $|y| \leq 2.8$ and a $\pt$ greater than or equal
to $80\GeV$. 
\item Jets are required to have $|y| \leq 2.8$ and $\pt\geq60\GeV$ in order to be counted.
\item A series of jet cleaning cuts were applied to eliminate various detector effects and suppress
beam and other non-collision backgrounds. 
Overall, these cuts reduce the available statistics by less than 0.1\%~\cite{njet2011}. 

\item Jets are only accepted if at least 70\%  
of their charged 
particle $\pt$ comes from the event vertex. 

\item Only events with a minimum of two selected jets are used in the analysis.
\squishend

\noindent For a total integrated luminosity of 2.43~pb$^{-1}$, 
approximately 500,000 multi-jet events survived the selection cuts. 

\section{Data Correction}\label{sec:unfolding}
A correction is needed to compare the measurements 
to theoretical predictions. 
The correction, which accounts for 
trigger inefficiencies, detector resolutions and other detector effects that affect the
jet counting, is performed in a single step using a bin-by-bin
unfolding method 
calculated from Monte Carlo simulations. For
each measured distribution, the corresponding Monte Carlo simulation cross
section using truth jets as defined in Section~\ref{sec:definitions} is
evaluated in the relevant bins, along with the equivalent
distributions obtained after the application of detector simulation and analysis
cuts. The ratio of the true to the simulated distributions
provides the multiplicative correction factor (unfolding factor) to be applied to the 
measured distributions. 

To perform the unfolding, 
ALPGEN+HERWIG/JIMMY AUET1 Monte Carlo simulation is used. The unfolding uncertainty is estimated taking into account several effects. 
One arises from the spread in correction factors coming from different generators
(ALPGEN+ HERWIG/JIMMY AUET1 and PYTHIA AMBT1).  A second detailed study is performed in which the simulated jet $\pt$, $y$ and $\phi$ resolution is varied.  Third, the shape of the simulated distributions is varied in order to account for possible biases 
caused by the input distributions. Samples with a trigger inefficiency in the crack
region, with different pile-up rejection cuts and different primary vertex multiplicity distributions are also used to 
estimate the uncertainty arising from trigger effects and from the impact of overlapping proton-proton
collisions.
All these effects impact the unfolding systematics, and their uncertainties are ultimately added in 
quadrature to provide the final systematic uncertainty in the unfolding correction. Statistical
uncertainties on the unfolding factors are important for certain bins and
added to the total uncertainty.

The combined systematic uncertainty on the data correction is less than 10\% in most bins except those affected by high statistical uncertainties where the uncertainty can be up to 20\%.

The systematic uncertainties in the luminosity calculation affect all cross section measurements,
but cancel out in all measurements where cross-section ratios are involved. The luminosity
of the dataset used in this paper has been calculated to be 2.43$\pm$0.08~pb$^{-1}$~\cite{lumi}
and the associated uncertainty is not shown in the figures. 

\section{Uncertainty on the Jet Energy Scale}

The jet energy scale uncertainty is the dominant uncertainty for
most results presented in this paper. The steeply falling cross sections 
as a function of jet
$\pt$ implies that even a relatively small uncertainty in the determination of the 
jet $\pt$ translates into a 
substantial change in the cross sections as events migrate along the steeply falling curve. 

The jet energy scale in ATLAS~\cite{atlas_jes_study} has been calculated
using jets from a dijet sample without near-by activity in the calorimeter.  
For a multi-jet analysis, a set of additional systematic uncertainties need to be considered.
These uncertainties arise from changes in the 
calorimeter response to jets of different flavors
as well as the presence of nearby activity in the calorimeter on the jet energy measurement.

In summary, the jet energy scale uncertainty is primarily
made of three components: the uncertainty calculated
for isolated jets, the uncertainty caused by the presence
of nearby calorimeter deposits, and the flavor composition
uncertainty. The uncertainty on the energy scale of
isolated jets is the largest contributor to the total uncertainty
in most bins, except for jets in the five and six-jet
bins and of $\pt < 200 $GeV, for which the flavor composition
uncertainty is comparable. The positive systematic
uncertainty on the jet energy scale of isolated jets falling
in the barrel and in high-multiplicity bins varies from 5$\%$
at 60 GeV to 2.5$\%$ at 1 TeV. In the three-jet and four-jet
bins, where the flavor composition is better constrained,
the systematic uncertainty is at most 3.5$\%$. The negative
systematic uncertainty is smaller and $\sim$3\% across all $\pt$
in the barrel. The impact of nearby calorimeter deposits
is small, increasing the overall uncertainty by at most 1$\%$.

\section{Results}
In this section, measurements\footnote{All measurements in this section have been compiled in Tables that can be found in HEPDATA.  The NLO pQCD calculation results are also presented in the Tables when applicable.} corrected to the particle level are compared to theoretical predictions. 
For comparisons to leading-order Monte Carlo simulations, the anti-$k_t$ algorithm with resolution parameter $R = 0.4$ 
is used to define a jet.  
In Figures~\ref{fig:njets}-3 and 5(b), 
the orange error band bracketing the measured cross section corresponds to the total 
systematic uncertainty, evaluated by adding the individual systematic uncertainties in quadrature 
but excluding the uncertainty coming from the luminosity measurement. 
The ratio of the predictions from the Monte Carlo simulations  
to the measurements is shown at the bottom of each figure. 
All Monte Carlo simulations are normalized to the measured inclusive two jet cross section, these factors are shown on the relevant figures.

Most ALPGEN Monte Carlo simulations studied predict an inclusive multi-jet cross section 
similar to the measured cross section, 
while the PYTHIA Monte Carlo simulation requires scaling factors
which differ significantly from unity.
The differences in the normalization factors between ALPGEN+PYTHIA
MC09$^{\prime}$ and ALPGEN+HERWIG/JIMMY AUET1 illustrate differences between PYTHIA 
and HERWIG/JIMMY and their interplay with the matrix-element and parton-shower 
matching implemented in ALPGEN.  The normalization factor for SHERPA is the closest to unity. 

Differences between the different ALPGEN tunes illustrate the large impact of the 
underlying event and parton shower tune on the leading-order predictions.
It has been hypothesized that the 
different parton shower models available in PYTHIA and 
HERWIG may, through the MLM-matching used in ALPGEN, have a significant effect on the 
overall prediction for the cross sections. 
This issue would benefit from additional studies, 
and perhaps dedicated tuning of ALPGEN+PYTHIA or ALPGEN+HERWIG.

Figure~\ref{fig:njets}(a) shows the results for the cross section as a function of the inclusive jet multiplicity.  
\begin{figure}[!b]
\setlength{\unitlength}{1mm}
 \centering
  \begin{picture}(200,70)(5,0)
\put(10,0){\epsfig{figure=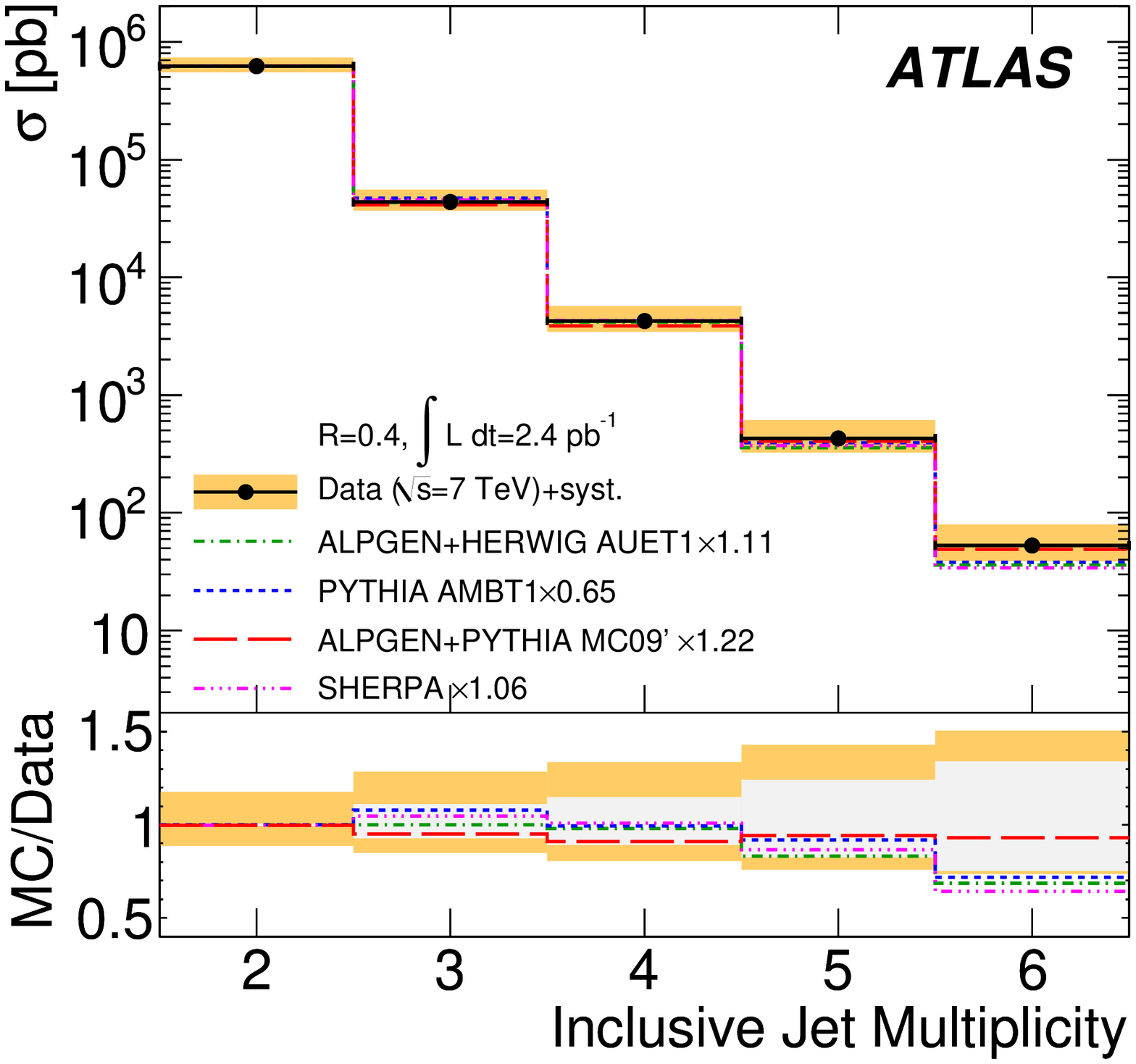,width=0.45\textwidth}}
\put(95,0){\epsfig{figure=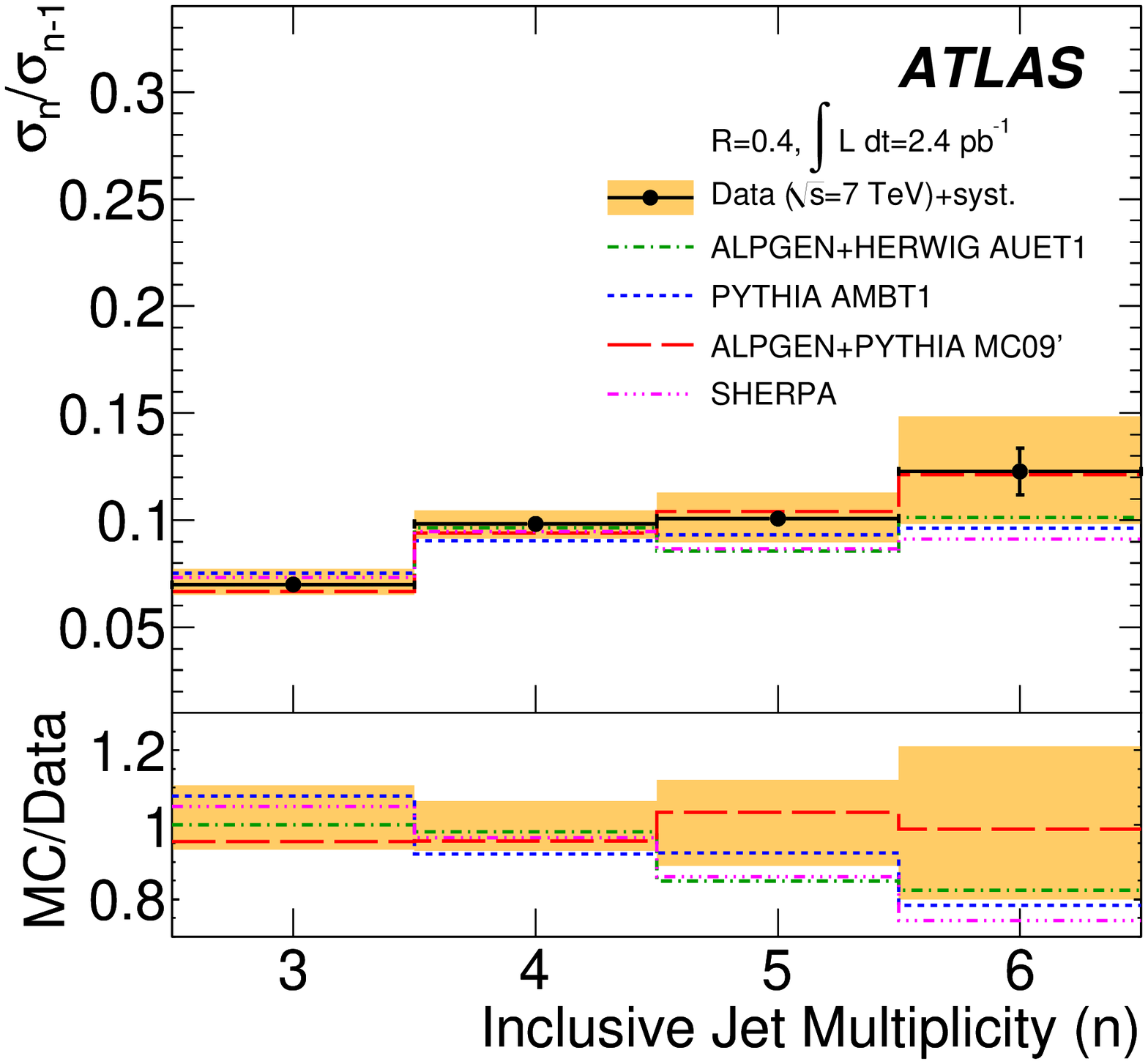,width=0.45\textwidth}}
 \put(55,65){$(a)$}\put(140,65){$(b)$}
\end{picture}
  \caption{Total inclusive jet cross section as a function of multiplicity (a). Ratio of the $n$-jet cross section to the $(n-1)$-jet cross section for values of $n$ 
   varying from three to six (b). \label{fig:njets}}   
   \end{figure}
The measurement systematics are dominated by the jet energy scale uncertainty and range from 10-20\% at 
low multiplicities to almost 30-40\% at high multiplicities. 
The Monte Carlo simulation predictions agree with the measured results across the full inclusive multiplicity
spectrum. 

A study that reduces significantly the impact of systematic uncertainties is the ratio of the $n$-jet 
to $(n-1)$-jet cross section as a function of multiplicity.  In this ratio, the impact of the 
jet energy scale uncertainty is significantly reduced and the uncertainty due to the luminosity 
cancels out. Figure~\ref{fig:njets}(b) shows the results for such a study.  

Both the data correction and the jet energy scale uncertainties contribute comparably to the total
systematic uncertainty, whereas the statistical uncertainties are 
smaller than the systematic uncertainties, and negligible in most bins.
All Monte Carlo simulations agree well with the data, yet there is a noticeable spread of $\sim$15-20\% in their predictions. 

The differential cross section for multi-jet events as a function of the jet $\pt$ 
is useful for characterizing the kinematic features.
Figure~\ref{fig:diffcrosssec} presents the $\pt$-dependent differential cross sections for the 
leading and third leading jet in multi-jet events.  
\begin{figure}[!htb]
\setlength{\unitlength}{1mm}
 \centering
  \begin{picture}(200,70)(5,0)
\put(10,0){\epsfig{figure=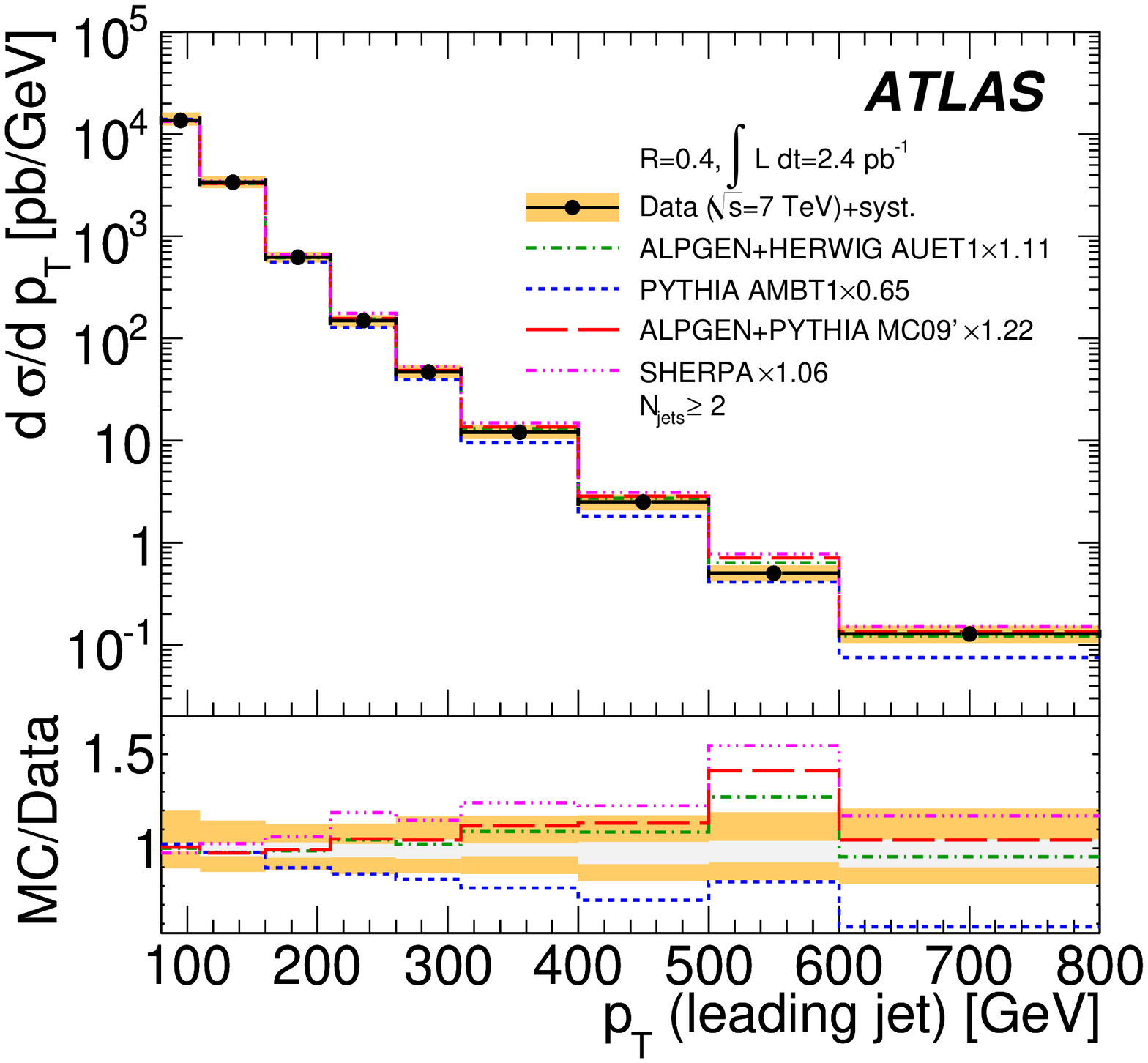,width=0.45\textwidth}}
\put(95,0){\epsfig{figure=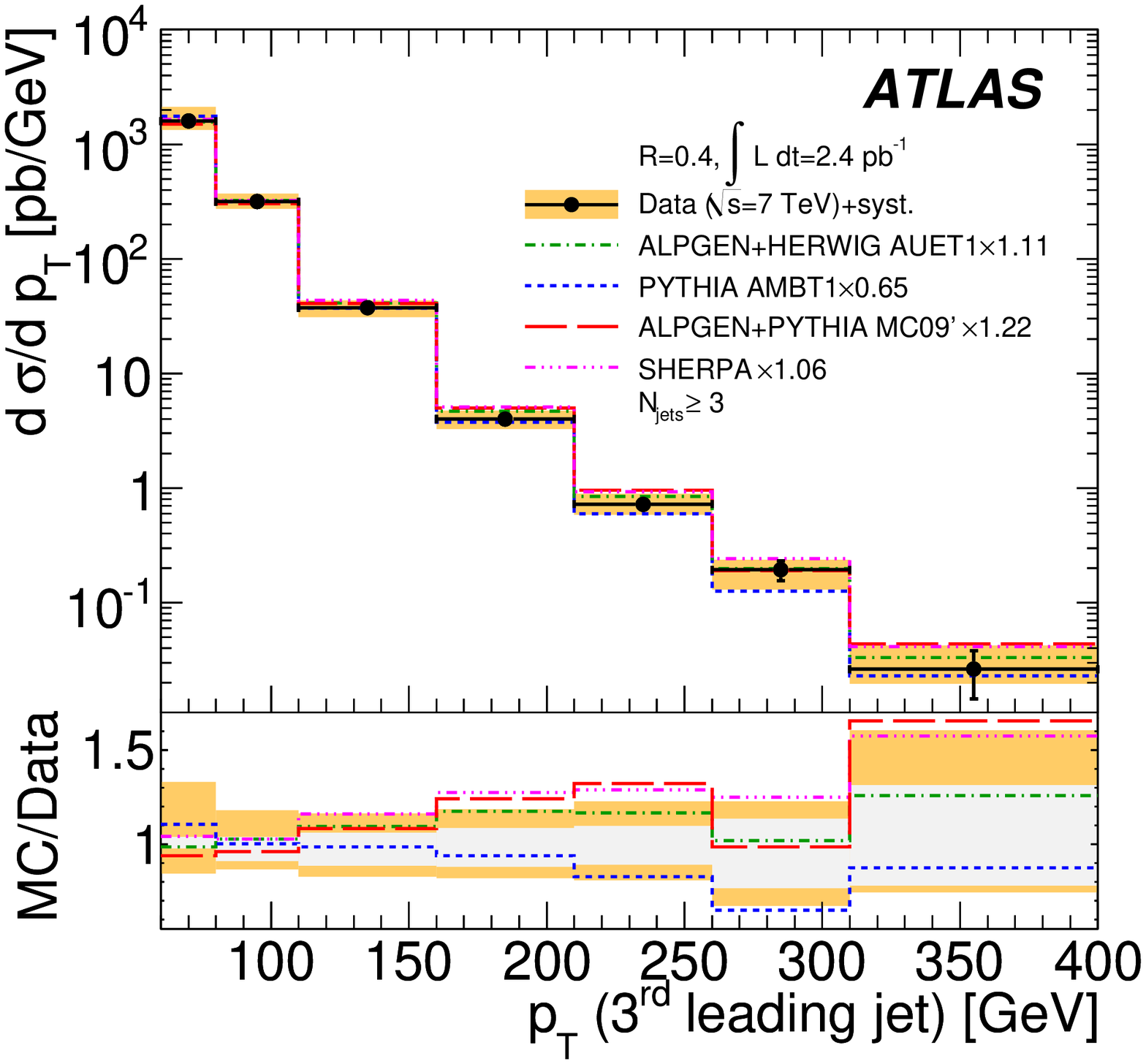,width=0.45\textwidth}}
 \put(55,65){$(a)$}\put(140,65){$(b)$}
\end{picture}

\caption{ Differential cross section as a function of  
leading jet $\pt$ for events with $N_{\rm jets}\geq 2$ (a), 
$3^{\rm rd}$ leading jet $\pt$ for events with $N_{\rm jets}\geq 3$ (b).} 
\label{fig:diffcrosssec}
\end{figure}
The systematic uncertainty in the measurement is 10-20\% across $\pt$ and increasing up
to 30\% for the fourth leading jet differential cross section.
The jet energy scale systematic uncertainty remains the dominant uncertainty in the measurement.
All Monte Carlo simulations
agree reasonably well with the data.  The PYTHIA AMBT1 Monte Carlo simulation predicts a somewhat steeper slope, compared to the data, as a function of the leading jet $\pt$ and of the second leading jet $\pt$.  The effect is most noticeable in the highest $\pt$ bins. 

\begin{figure}[!htb]
\setlength{\unitlength}{1mm}
 \centering
  \begin{picture}(200,50)(5,0)
\put(5,0){\epsfig{figure=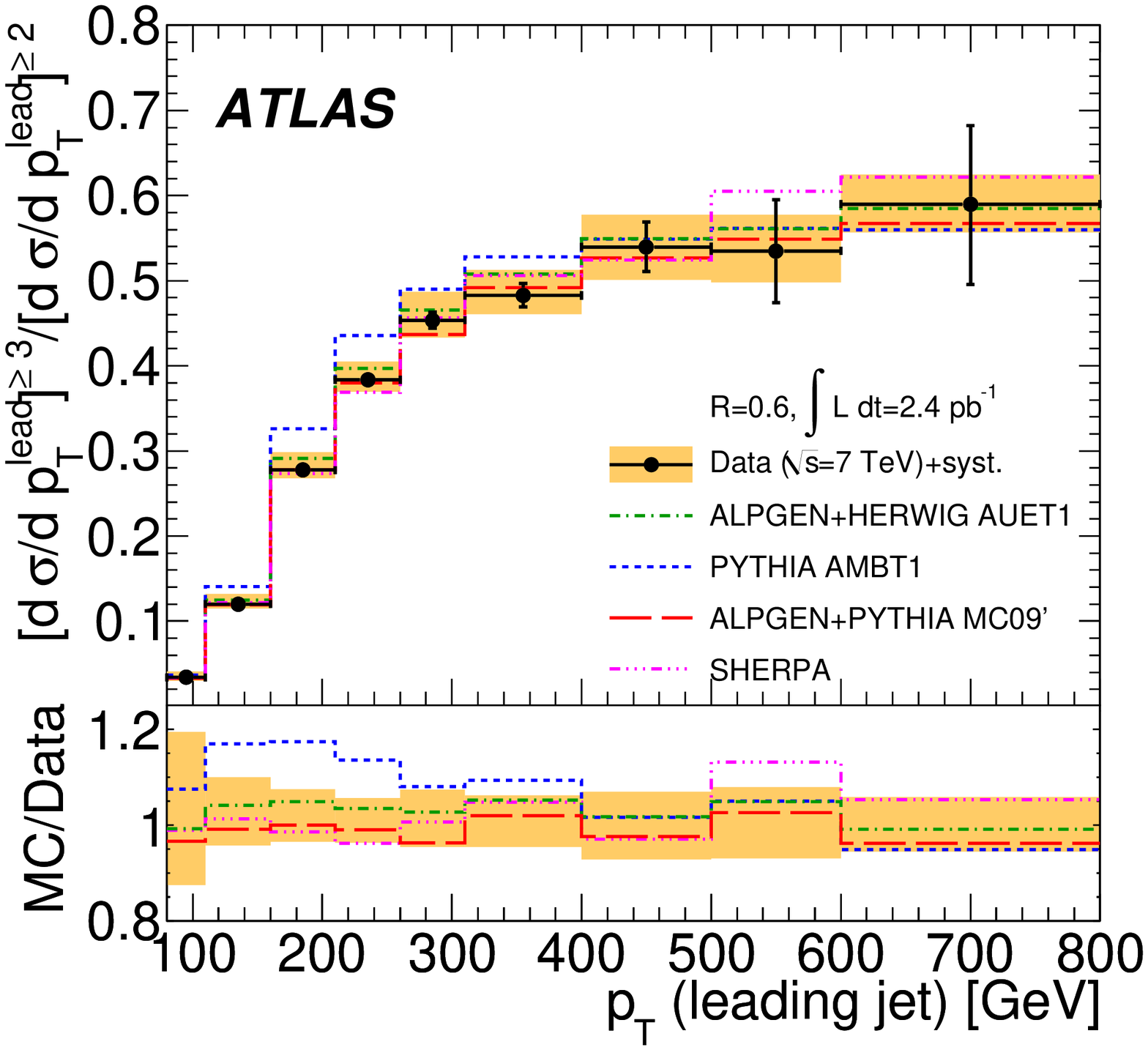,width=0.32\textwidth}}
\put(62.5,0){\epsfig{figure=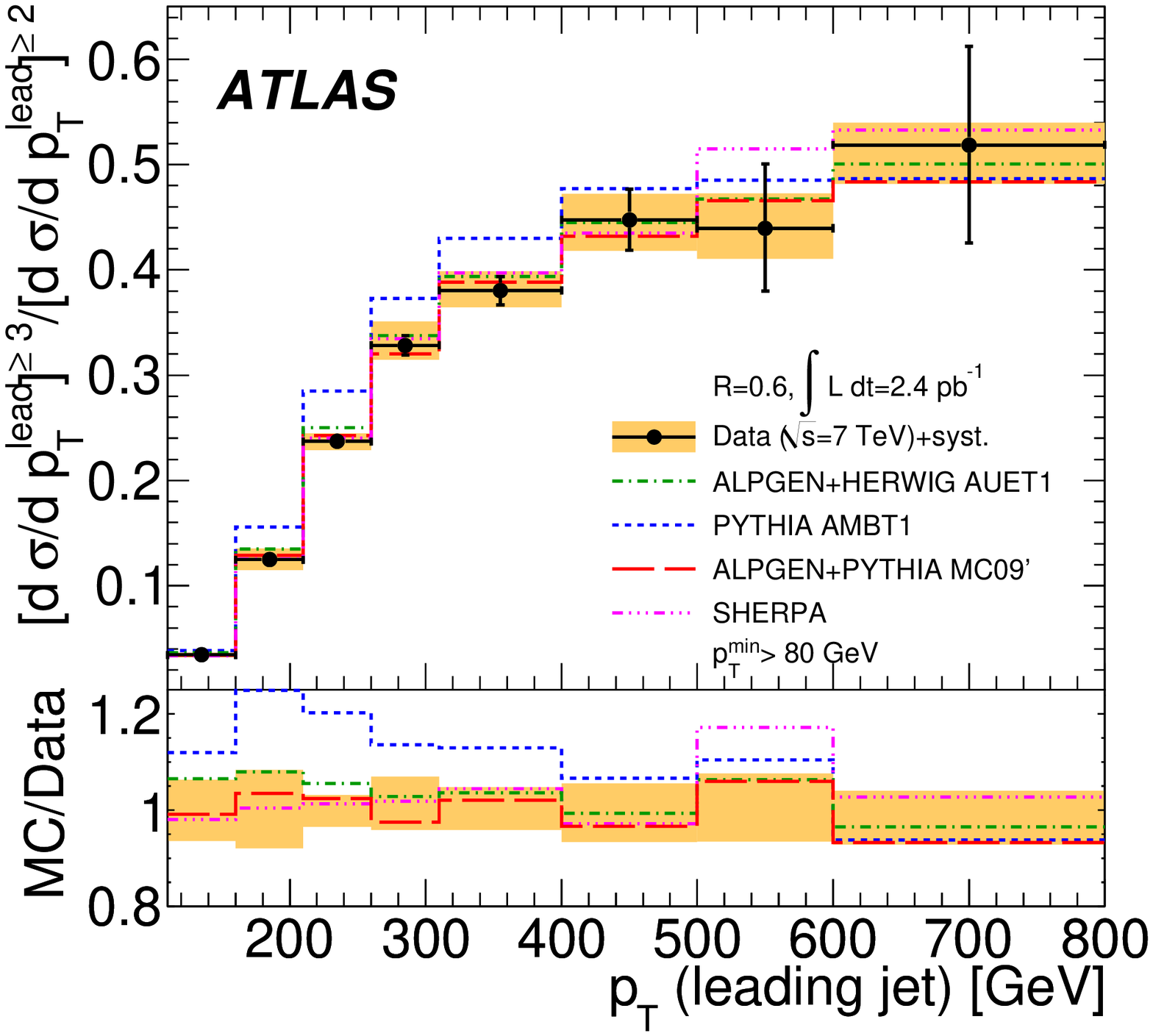,width=0.32\textwidth}}
\put(120,0){\epsfig{figure=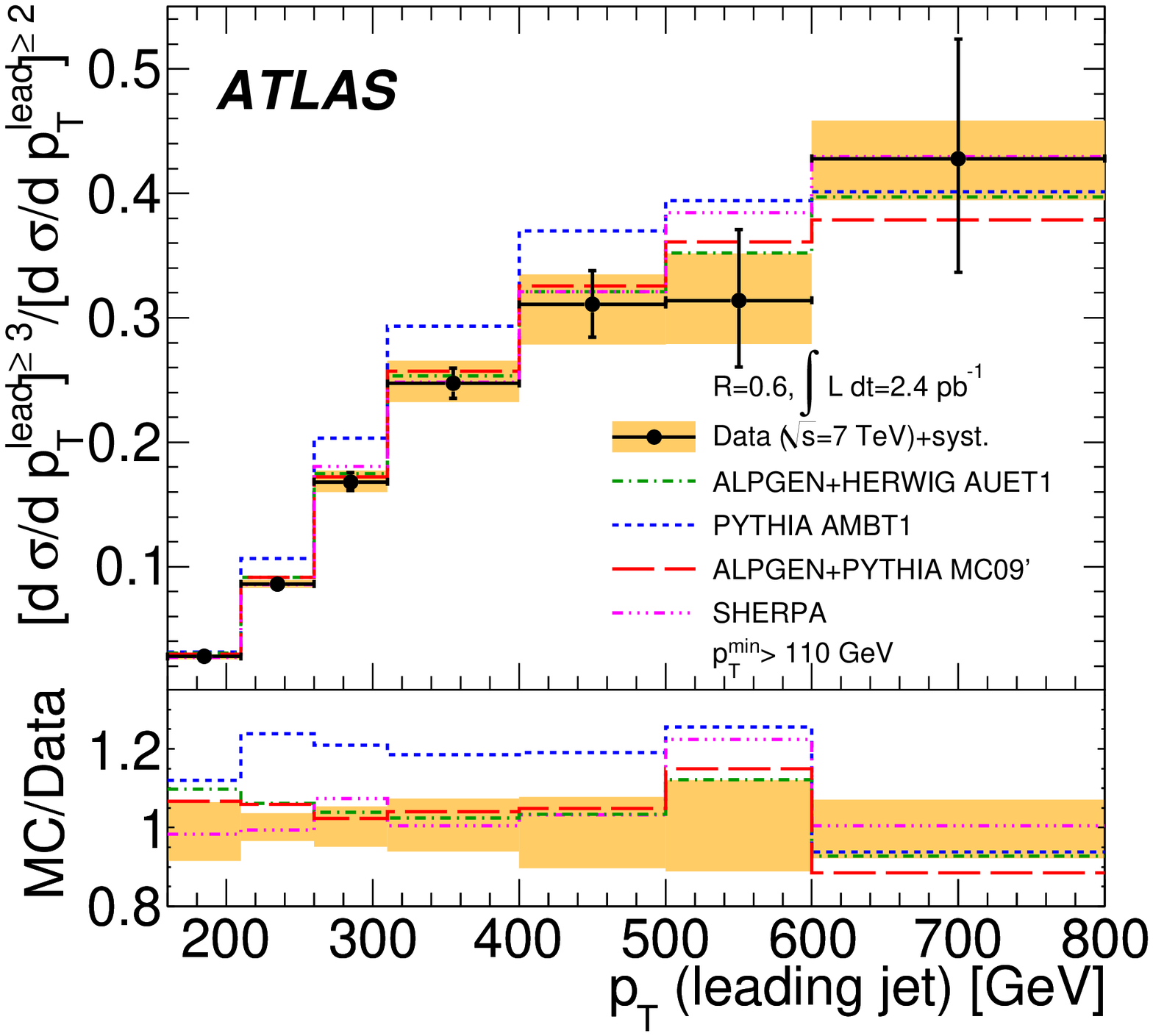,width=0.32\textwidth}}
\put(26,45){$(a)$}\put(83.5,45){$(b)$}\put(141,45){$(c)$}
\end{picture}

\caption{Three-to-two-jet differential cross-section ratio as a function of the leading jet $\pt$.  In the figures, a resolution parameter $R=0.6$ is used.  The three figures contain a minimum $\pt$ cut for all non-leading jets of (a) 60$\GeV$ (b) 80$\GeV$ and (c) 110$\GeV$.
}
\label{fig:threetotwoLO_lead}
\end{figure}

\begin{figure}[!htb]
\setlength{\unitlength}{1mm}
 \centering
  \begin{picture}(200,50)(5,0)
\put(5,0){\epsfig{figure=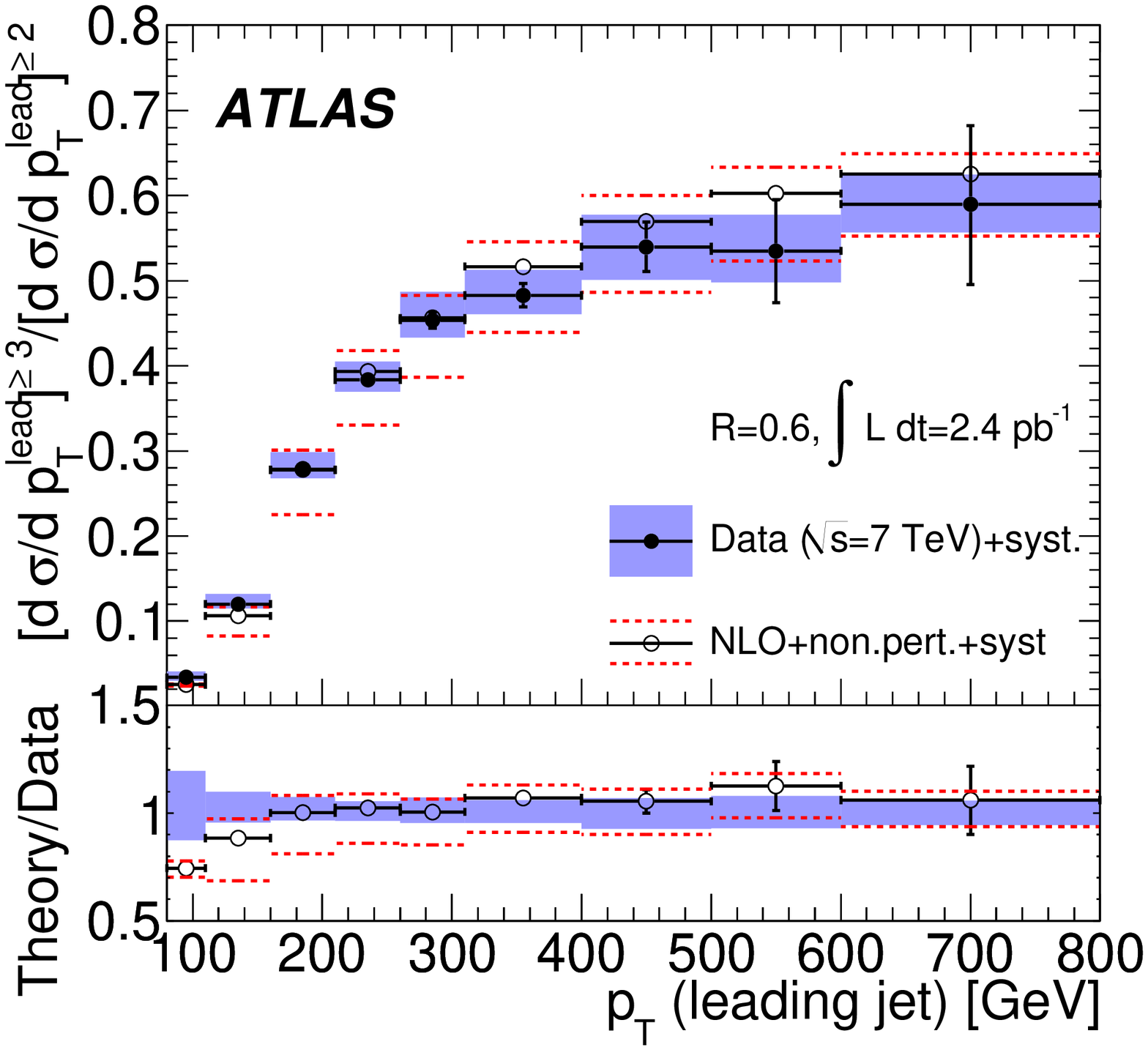,width=0.32\textwidth}}
\put(62.5,0){\epsfig{figure=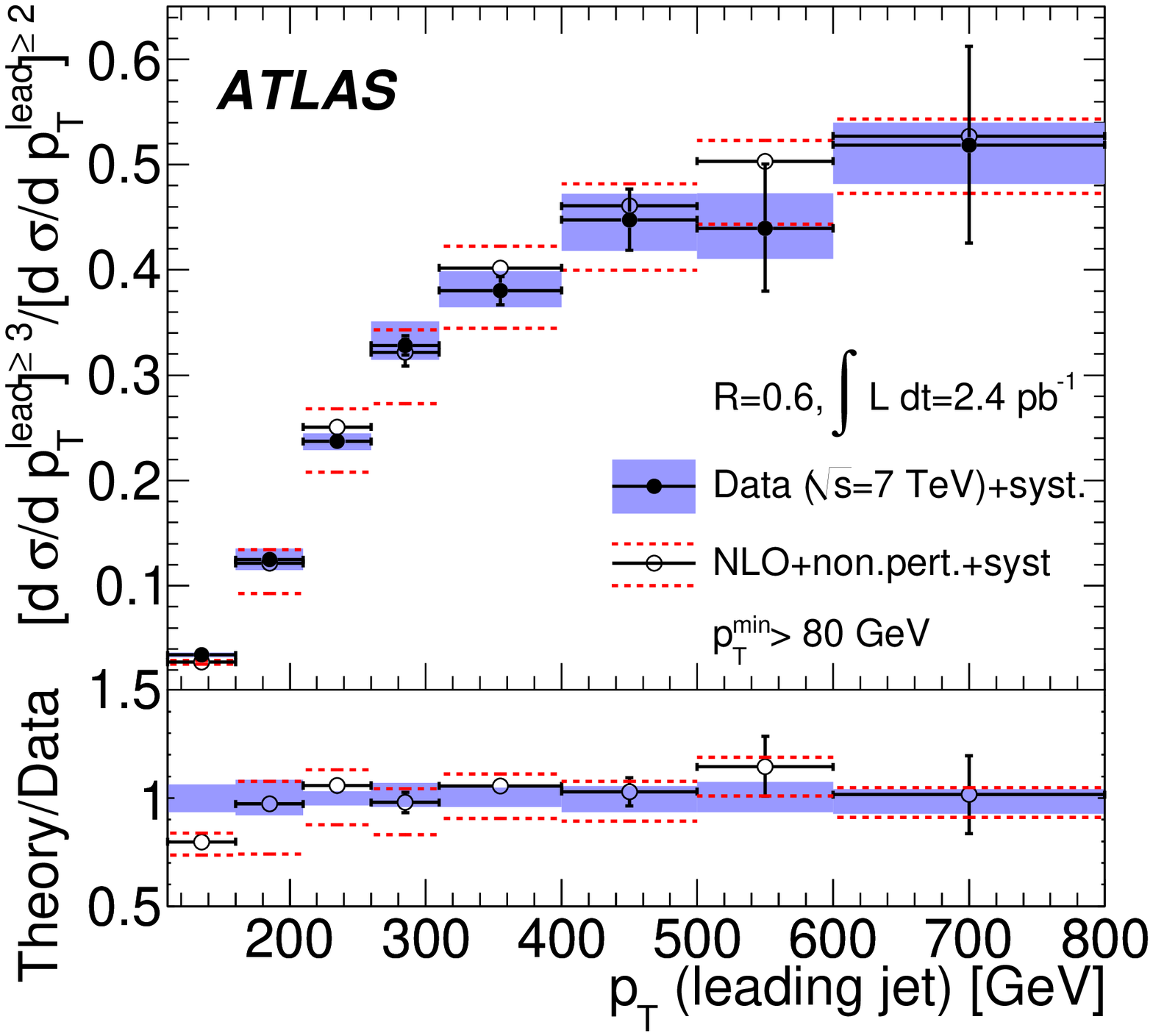,width=0.32\textwidth}}
\put(120,0){\epsfig{figure=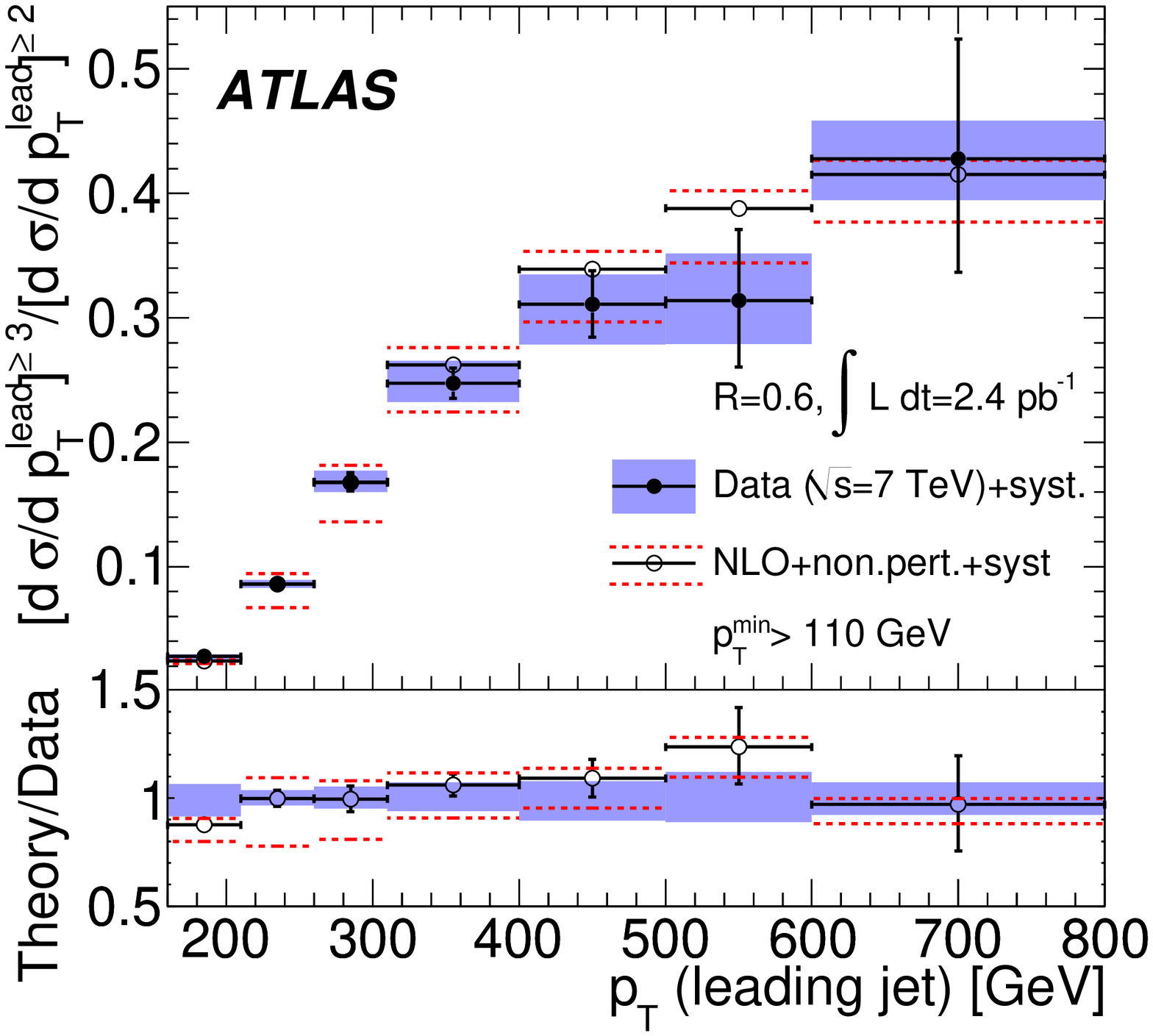,width=0.32\textwidth}} 
 \put(26,45){$(a)$}\put(83.5,45){$(b)$}\put(141,45){$(c)$}
\end{picture}
\caption{
Three-to-two-jet differential cross-section ratio as a function of the leading jet $\pt$.  
In the figures a resolution parameter $R=0.6$ is used.  The three figures contain a minimum $\pt$ cut for all non-leading jets of (a) 60$\GeV$ (b) 80$\GeV$ and (c) 110$\GeV$.
The results are 
compared to a NLO pQCD calculation with the \textsf{MSTW 2008 nlo} pdf set .
  The systematic uncertainties on the theoretical prediction are 
 shown as dotted red lines above and below the theoretical prediction. 
 \label{fig:threetotwoNLO_lead}
}
\end{figure}

\begin{figure}[!htb]
 \setlength{\unitlength}{1mm}
 \centering
  \begin{picture}(200,70)(5,0)
 \put(10,0){\epsfig{figure=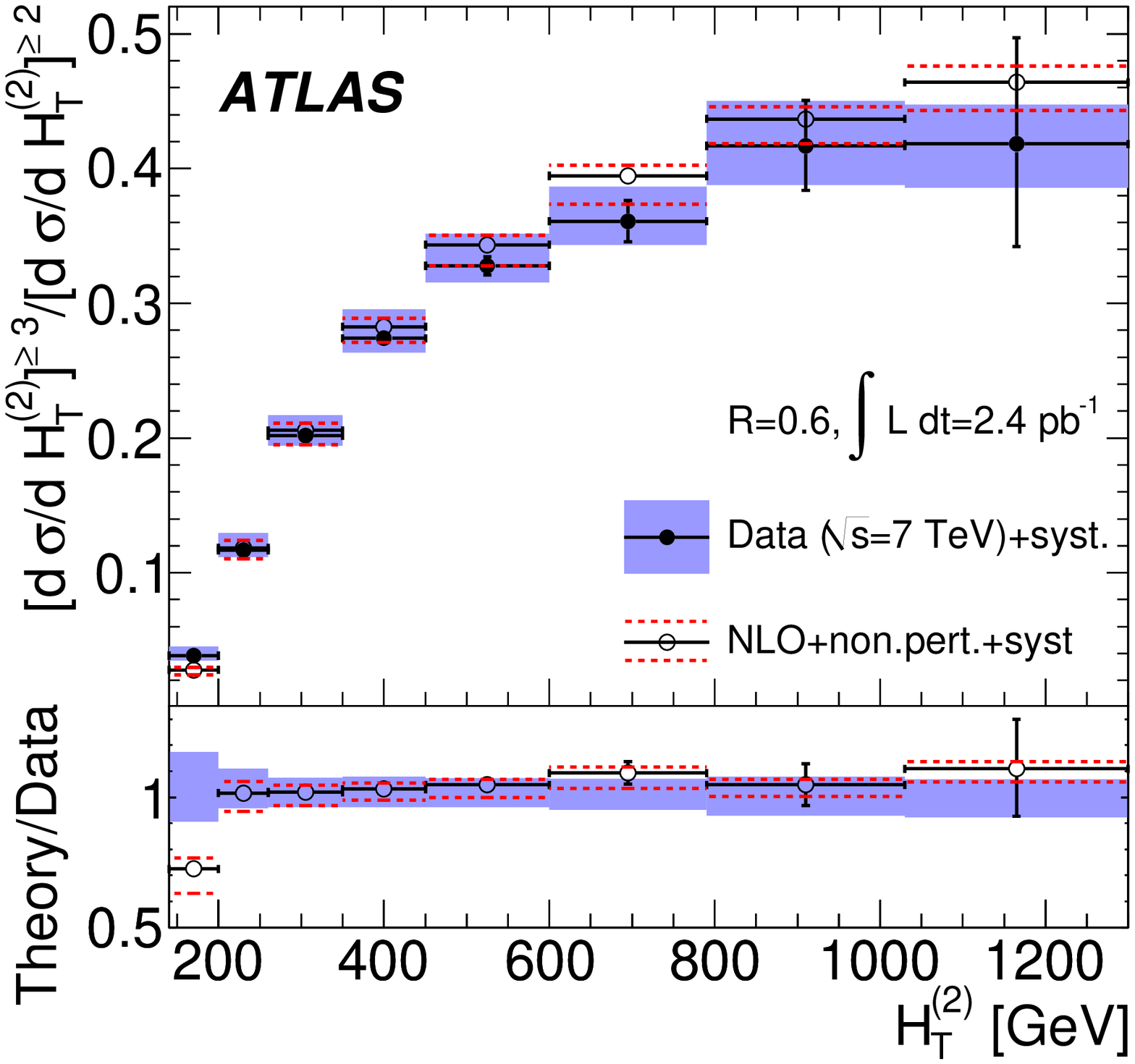,width=0.45\textwidth}}
  \put(95,0){\epsfig{figure=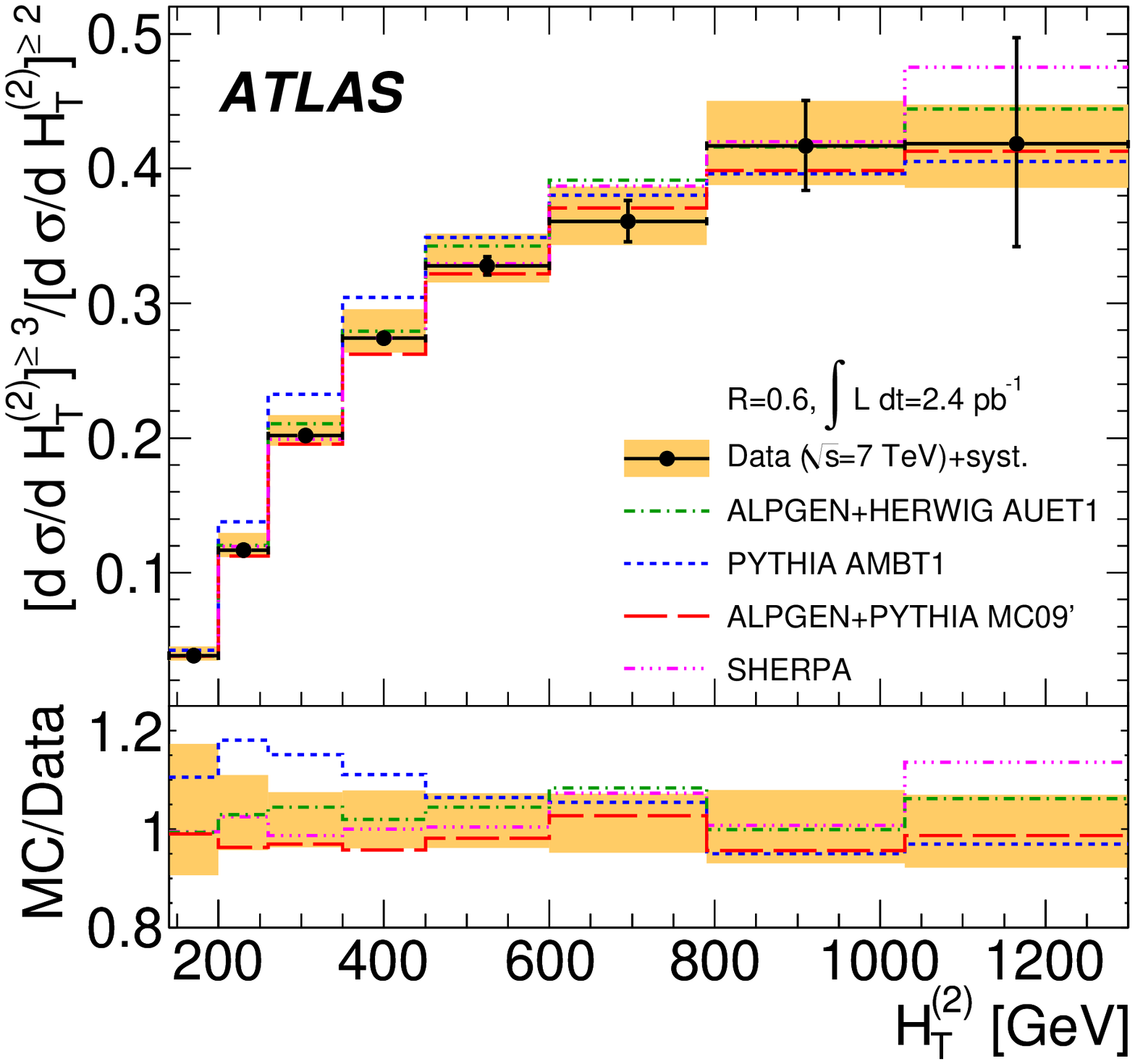,width=0.45\textwidth}}
   \put(40,63){$(a)$}\put(125,63){$(b)$}
\end{picture}
      \caption{Three-to-two-jet differential cross-section ratio as a function of the sum of the $\pt$ 
   of the two leading jets using $R=0.6$.  
  The left and right figures present the same measurements and error bands.
 The data are compared to (a) a NLO pQCD calculation and (b) several leading-order Monte Carlo simulations.
  The systematic uncertainties on the theoretical prediction for the NLO pQCD calculations are 
   shown as dotted red lines above and below the theoretical prediction.  
 }
   \label{fig:threetotwoNLO_HT}
\end{figure}
A measurement with particular sensitivity to limitations in the leading-order Monte Carlo simulations and NLO pQCD calculations is the ratio of the inclusive three-to-two-jet differential cross section as a function of some characteristic scale in the event.  In this measurement, the uncertainty in the luminosity determination cancels out, uncertainties in the jet energy scale are reduced, and statistical uncertainties are limited only by the inclusive three-jet sample.

For tuning Monte Carlo simulations, the three-to-two-jet ratio as a function of the leading jet $\pt$ has been shown to be particularly sensitive to effects from final state radiation~\cite{Sawyer}.   Figure \ref{fig:threetotwoLO_lead} presents the results on the measurement of the three-to-two-jet cross section ratio as a function of leading jet $\pt$ for jets built with the anti-$k_t$ algorithm using the resolution parameter R=0.6 and with different minimum $\pt$ cuts for all non-leading jets\footnote{Results (not shown) were also obtained using $R=0.4$ and are compiled in Tables in HEPDATA.}.
The systematic uncertainties on the measurement are small ($\sim$5\%), except in the lowest $\pt$ bin, where uncertainties in the unfolding correction dominate. 
ALPGEN+HERWIG AUET1 and ALPGEN+PYTHIA MC09$^{\prime}$ describe the data well, and the agreements appear to be largely independent of the tunes chosen.  
SHERPA also describes the data well.  
PYTHIA AMBT1 predicts a higher ratio than what is measured over the $\pt$ range from 200$\GeV$ to 600$\GeV$\footnote{Further results for more Monte Carlo simulations are compiled in Tables in HEPDATA. The disagreement observed when PYTHIA Perugia2010, PYTHIA MC09 and HERWIG++ are used is similar to that observed here.}. 
The systematic uncertainty in the lowest $\pt$ bin decreases significantly as the minimum $\pt$ cut is raised to 80$\GeV$ for all jets. 

Figure \ref{fig:threetotwoNLO_lead} presents the same measurement results as Figure \ref{fig:threetotwoLO_lead}, except the data are now compared to the NLO pQCD calculations corrected for non-perturbative effects. The MSTW 2008 nlo pdf set has been used, but comparable results are obtained with the CTEQ 6.6 pdf set. The systematic uncertainties on the theoretical predictions are shown as dotted red lines above and below the theoretical prediction. The NLO pQCD calculations describe the data well, except in the lowest $\pt$ bin, where there is a large discrepancy.  The discrepancy diminishes significantly once the minimum $\pt$ for all jets is raised to 110$\GeV$ and the $\pt$ of the leading jet is required to be greater than 160$\GeV$.  

Additional NLO pQCD calculations of the three-to-two-jet cross section ratio were performed as a function of different kinematic variables, such as $\HT$, the sum of the $\pt$ of the two leading jets ($H_T^{(2)}$) and the sum of the $\pt$ of the three leading jets.  The NLO pQCD calculation for the ratio as a function of $H_T^{(2)}$ was found to give the smallest theoretical scale uncertainty and is, therefore, most sensitive to input parameters such as $\alpha_S$.
Figure~\ref{fig:threetotwoNLO_HT} shows a comparison of the measurement to both NLO pQCD (a) and leading order calculations (b) for R = 0.6.
 Scale uncertainties of the NLO pQCD calculations
 are larger for jets with $R=0.4$ than with $R=0.6$.
The theoretical uncertainty of the NLO pQCD calculations shown in Figure~\ref{fig:threetotwoNLO_lead}  is comparable to the measurement uncertainties, but is significantly reduced compared to theoretical uncertainties presented in Figure~\ref{fig:threetotwoNLO_HT}.  With the reduced theoretical uncertainty, the disagreement between data and the NLO pQCD calculations in the lowest $H_T^{(2)}$ bin is now enhanced. Due to the kinematic cuts applied in the analysis, the NLO pQCD calculations only account for the lowest-order contribution to the two-jet cross section in the region where the sum of the first and second leading jet $\pT$ is less than 160$\GeV$.  Consequentially, this effective leading-order estimation is subject to large theoretical uncertainties, which might be responsible for the observed discrepancy.   

A comparison of the same measurement to leading-order Monte Carlo simulations is given in Figure~\ref{fig:threetotwoNLO_HT} (b). The general agreement between leading-order Monte Carlo simulations with the measurements follows the same general trends as the comparison of the three-to-two-jet ratio versus leading jet $\pt$ shown in Figure~\ref{fig:threetotwoNLO_lead}.

\section{Summary and Conclusion}
A first dedicated study of multi-jet events has been performed using the ATLAS detector at a 
center-of-mass energy of $7\TeV$ with an integrated luminosity of 2.43~pb$^{-1}$.  For events containing two 
or more jets with $\pt\geq60\GeV$, of which at least one has $\pt\geq80\GeV$, good agreement is 
found between data and leading-order Monte Carlo simulations with 
parton-shower tunes that describe 
adequately the ATLAS $\sqrt{s}=7\TeV$ underlying-event data. The agreement is found
after the predictions of the Monte Carlo simulations are normalized to the measured
inclusive two-jet cross section. 
The models have been 
compared to multi-jet inclusive and differential 
cross sections.
The present study extends up to a
multiplicity of six jets, up to jet $\pt$ of $800\GeV$ and up to event $\HT$ 
of 1.6$\TeV$.   

All models reproduce the main features of the multijet data.
ALPGEN, 
which contains higher multiplicity tree-level matrix elements matched to parton showers, generally 
describes the shapes well, whether the HERWIG or PYTHIA parton showers are used.  
Similarly, SHERPA describes the data well.
However, PYTHIA, which contains a $2\rightarrow 2$ leading-order matrix element 
augmented by parton showers, 
shows only a marginal agreement with the data.
This is most noticeable 
in the three-to-two-jet cross section ratios.

A measurement of the three-to-two-jet cross section ratio as a function of the leading jet $\pt$ and
the sum of the two leading jet $\pt$s has also been performed and is described by 
ALPGEN, SHERPA and a NLO pQCD 
calculation, albeit with a significant discrepancy in the lowest $\pt$ bin for the latter comparison. Comparisons with 
NLO pQCD calculations may be useful for constraining parameters, such as parton distribution
functions or 
the value of the strong coupling constant, $\alpha_S$, given
that the systematic uncertainties from the measurement are comparable to the theoretical uncertainties.





\bigskip 
\bibliography{multijet-paper}

\providecommand{\href}[2]{#2}\begingroup\raggedright\begin{thebibliography}{10}

\bibitem{multijetpaper}
{The ATLAS Collaboration}, CERN-PH-EP-2011-098  (2011).

\bibitem{atlas_detector_paper}
{The ATLAS Collaboration},
\href{http://dx.doi.org/10.1088/1748-0221/3/08/S08003}{JINST 3 (2008)\,S08003}.

\bibitem{atlas_trigger08}
R.~Achenbach et al., JINST 3 (2008)\,P03001.

\bibitem{Cacciari_antikt}
M.~Cacciari, G.~P. Salam, and G.~Soyez, JHEP 04 (2008)063.

\bibitem{Fastjet}
M.~Cacciari, G.~P.~Salam, G.~Soyez, http://fastjet.fr/.

\bibitem{UE}
A.~Moraes, C.~Buttar, and I.~Dawson, Eur. Phys. J. C50 (2007)435.

\bibitem{Frixione}
{S. Frixione and G. Ridolfi}, Nucl. Phys. B507  (1997)315.

\bibitem{Mangano:2002ea}
M.~L. Mangano et al., JHEP 07 (2003)001.

\bibitem{Pumplin:2002vw}
J.~Pumplin et al., JHEP 07 (2002)012.

\bibitem{Sjostrand:2007gs}
T.~Sjostrand, S.~Mrenna, and P.~Z. Skands, Comput. Phys. Commun. 178 (2008)852.

\bibitem{Sjostrand:2000wi}
T.~Sjostrand et al., Comput. Phys. Commun. 135 (2001)238.

\bibitem{Corcella:2000bw}
G.~Corcella et al., JHEP 01 (2001)010.

\bibitem{Corcella:2002jc}
G.~Corcella et al.,
\href{http://arxiv.org/abs/hep-ph/0210213}{arXiv:hep-ph/0210213}.

\bibitem{Butterworth:1993ig}
J.~M. Butterworth and J.~R. Forshaw,
J. Phys. G19 (1993)1657.

\bibitem{Butterworth:1996zw}
J.~M. Butterworth, J.~R. Forshaw, and M.~H. Seymour, Z. Phys. C72 (1996)637.

\bibitem{ATL-PHYS-PUB-2010-002}
{The ATLAS Collaboration}, {ATLAS Note} ATLAS-PHYS-PUB-2010-002 (2010).

\bibitem{ATL-PHYS-PUB-2010-014}
{The ATLAS Collaboration}, {ATLAS Note} ATLAS-PHYS-PUB-2010-014 (2010).

\bibitem{Sherpa}
{T. Gleisberg et al.}, JHEP 02 (2009)007.

\bibitem{Sherstnev:2008dm}
A.~Sherstnev and R.~S. Thorne,
\href{http://arxiv.org/abs/0807.2132}{arXiv:0807.2132 [hep-ph]}.

\bibitem{Martin:2009iq}
A.~D. Martin et al., Eur. Phys. J. C63 (2009)189.

\bibitem{SimJINST}
{The ATLAS Collaboration}, Eur. Phys. J. C70 (2010)823,
  \href{http://arxiv.org/abs/1005.4568}{arXiv:1005.4568}.

\bibitem{geant}
S.~Agostinelli et al., Nucl. Instr. and Meth. A 506 (2003)250.

\bibitem{NLOJet}
{Z. Nagy}, Phys. Rev. D68 (2003){094002}.

\bibitem{Nadolsky:2008zw}
P.~M. Nadolsky et al.,
  \href{http://dx.doi.org/10.1103/PhysRevD.78.013004}{Phys. Rev. D78
  (2008)013004},
\href{http://arxiv.org/abs/0802.0007}{arXiv:0802.0007 [hep-ph]}.

\bibitem{Bahr:2008pv}
M.~Bahr, S.~Gieseke, M.~Gigg, D.~Grellscheid, K.~Hamilton, et al.,
  \href{http://dx.doi.org/10.1140/epjc/s10052-008-0798-9}{Eur.Phys.J. C58
  (2008)639--707}, \href{http://arxiv.org/abs/0803.0883}{arXiv:0803.0883
  [hep-ph]}.

\bibitem{ATLAS-CONF-2010-027}
{The ATLAS Collaboration}, ATLAS Note ATLAS-CONF-2010-027 (2010).

\bibitem{ATLAS-CONF-2010-069}
{The ATLAS Collaboration}, {ATLAS Note} ATLAS-CONF-2010-069 (2010).

\bibitem{:2010wv}
{Atlas Collaboration} Collaboration, G.~Aad et al.,
  \href{http://dx.doi.org/10.1140/epjc/s10052-010-1512-2}{Eur.Phys.J. C71
  (2011)1512}, \href{http://arxiv.org/abs/1009.5908}{arXiv:1009.5908 [hep-ex]}.

\bibitem{atlas_jes_study}
{The ATLAS Collaboration}, {ATLAS Note} ATLAS-CONF-2011-032 (2011).

\bibitem{njet2011}
{The ATLAS Collaboration}, ATLAS Note ATLAS-CONF-2011-043 (2011).

\bibitem{lumi}
{The ATLAS Collaboration}, ATLAS Note ATLAS-CONF-2011-011 (2011).

\bibitem{Sawyer}
L.~Sawyer,
PoS DIS2010 (2010)135.

\end{thebibliography}\endgroup

\end{document}